\documentclass[ showpacs,showkeys,preprintnumbers,amsmath,amssymb,floatfix,prd]{revtex4}

\usepackage{graphicx}
\usepackage{dcolumn}
\usepackage{bm}

\newcommand{\beq}{\begin{equation}}
\newcommand{\eeq}{\end{equation}}
\newcommand{\bey}{\begin{eqnarray}}
\newcommand{\eey}{\end{eqnarray}}

\begin{document}                             

\title{ A bridge between unified cosmic history by  $f(R)$-gravity and  BIonic system}

\author{Alireza Sepehri $^{1}$\footnote{alireza.sepehri@uk.ac.ir}, Salvatore  Capozziello $^{2,3,4}$\footnote{capozziello@na.infn.it}, Mohammad Reza Setare $^{5}$\footnote{rezakord@ipm.ir},  }
\address{$^1$ Faculty of Physics, Shahid Bahonar University, P.O. Box 76175, Kerman, Iran.\\$^{2 }$ Dipartimento
di Fisica, Universit´a di Napoli "Federico II", I-80126 - Napoli,
Italy. \\
$^{ 3 }$ INFN Sez. di Napoli, Compl. Univ. di Monte S. Angelo,
Edificio G, I-80126 - Napoli, Italy,\\ $^{4}$ Gran Sasso
Science Institute (INFN), Viale F. Crispi, 7, I-67100, L'Aquila,
Italy \\ $^5$  Department of Science, Campus of Bijar,
University of Kurdistan, Bijar, Iran.}

\begin{abstract}
Recently, the cosmological deceleration-acceleration transition
redshift in $f(R)$ gravity has been considered in order to   address consistently   the problem of cosmic evolution. It is possible to show that the
deceleration parameter  changes sign at a given redshift
according to  observational data. Furthermore, a  $f(R)$ gravity
cosmological model can be constructed in brane-antibrane system starting from  the very early universe and accounting for  the cosmological redshift at  all phases of cosmic history,  from inflation to late time acceleration. Here we propose a $f(R)$ model where
 transition redshifts  correspond  to 
inflation-deceleration and  deceleration-late time
acceleration transitions starting froma BIon system.  At the  point where  the universe was
born, due to the transition of  $k$ black fundamental strings to the BIon
configuration, the redshift is approximately infinity and
decreases with reducing temperature ($z\sim T^{2}$). The BIon is a
configuration in flat space of a universe-brane and a parallel
anti-universe-brane connected by a wormhole. This wormhole is a
channel for flowing energy from extra dimensions into our
universe, occurring at inflation and decreasing with redshift  as $z\sim
T^{4+1/7}$. Dynamics consists with the fact that the wormhole misses its energy and  vanishes as soon as 
inflation ends and deceleration begins. Approaching two universe
branes together, a tachyon is originated, it grows up  and causes the formation of
a wormhole. We show that, in the framework of $f(R)$ gravity, the  cosmological redshift depends on the
tachyonic potential and has a significant decrease at
deceleration-late time acceleration transition point ($z\sim
T^{2/3}$). As soon as  today acceleration approaches, the redshift tends to
zero and the cosmological model reduces to the standard $\Lambda$CDM cosmology.


\end{abstract}
\date{\today}

 \maketitle
\section{Introduction}
Today cosmology is  is facing with two fundamental questions:
what is inflation and what is late time acceleration? In other
words, why and how do both the very early and the very late
universe expand with acceleration? Extended theories of gravity, and in particular $f(R)$ gravity seems a reliable approach to reply these questions and unifying the early time
inflation with late time acceleration era 
\cite{m1,m2,m3,m4,m5,annalen}. This model allows, in principle,  to study the inflationary
epoch, the transition to the non-phantom standard cosmology
(radiation/matter dominated eras) and today observed dark energy
epoch. For example,  the  evolution of
quintessence/phantom dominated era in modified $f(R)$ gravity has been widely
considered starting from very elementary models \cite{quintessence} up to more realistic approaches \cite{alcaniz}. 
In  Refs. \cite{m1,annalen}, it is demonstrated that this type of gravity unifies
the early-time inflation with late-time acceleration and can be
consistent with observational data. In another scenario,
authors unified $R^{m}$ inflation with late-time $\Lambda$CDM
epoch in the modified $f(R)$ gravity. They showed  that this model
is consistent with local tests like Newton law, stability of
Earth-like gravitational solution and  very heavy mass for
additional scalar degree of freedom \cite{m2}. Furthermore,
 a unified description of early-time inflation and
late-time acceleration has been studied in a non-linear modified $f(R)$
Horava-Lifshitz gravity. It has been demonstrated that the cosmological
dynamics details are generically different from those of other
viable $f(R)$ models \cite{m3}. Finally, it has been  shown
that inflation and the late-time acceleration  can be realized in
a modified Maxwell-$f(R)$ gravity which is consistent with Solar
System tests \cite{m4}.

Furthermore,  a class of $f(R)$ gravity models has been analyzed  considering
the Ricci scalar as a function of the redshift $z$. In this case the $f(R)$ gravity model  reduces to
$\Lambda$CDM at $z \simeq 0$. The observational viability of this
class has been examined and the deceleration-acceleration
transition redshift has been constrained by suitable sets  of observational  data
\cite{m5}. However, an important  question is on the origin
of unified model and transition redshift in $f(R)$ gravity.  In other words,  the question is if there is a fundamental origin of reliable $f(R)$ gravity models that can  be tested  by observations. 

A possible
answer to this question can come  from a  brane-antibrane system. In fact, the
evolution of this system can change the cosmological redshift and
produce different epochs of cosmic history (see also \cite{mariam} for a cosmographic analysis).  When the $k$ black
fundamental strings transited to the BIon configuration, our observed universe
is born and the redshift is approximately infinity ($z\sim \infty$).
The BIon is a configuration in flat space of a universe-brane and
a parallel anti-universe-brane connected by a wormhole
\cite{m6,m7}. At this stage, large amount of energy is
transferred from anti-brane to our own universe brane: consequently the redshift
reduces and inflation occurs. At second stage, the wormhole dies  and
there is no channel for flowing energy into our universe. In
this condition, inflation ends  and the cosmological redshift
results reduced with lower velocity. With decreasing  separation
distance between the  two universe branes, a tachyon is originated. It  grows and
causes the formation of another wormhole. At this stage, the late time
acceleration starts and the cosmological redshift accelerates to
lower values.

 The outline of the
paper is as the following.  In Sect. \ref{o1}, we will discuss a
bridge between  the inflationary stage in BIon system and $f(R)$
gravity. We will show that the redshift parameter in the $f(R)$ model
depends on the evolution of the  wormhole in BIon. In Sect. \ref{o2},
we study the behavior of redshift in the second stage when the
wormhole dies and  there is only a pair of universe brane and
antibrane. In Sect. \ref{o3}, we obtain the amount of redshift
in third stage where a  new tachyonic wormhole is formed between
branes and accelerates the destruction of the two universes. In Sect.
\ref{o4}, we  discuss  the results and draw   the conclusion.

\section{ Stage 1: The cosmological redshift in the early time inflation}
\label{o1}

Let us  assume that there is only a  $k$ black
fundamental string at the beginning of  time. In our model, the universe
is  born at the  point where the thermodynamics of $k$
non-extremal black fundamental strings is matched to that of
the BIon configuration. We will construct our $f(R)$ gravity model in BIon and
discuss that  cosmological redshift depends on the number of
branes and the distance between them.

The supergravity solution for the $k$ coincident non-extremal black
F-strings lying along the z direction is:
\begin{eqnarray}
&& ds^{2} = H^{-1}(-f dt^{2} + dz^{2})+ f^{-1}dr^{2} + r^{2}d\Omega_{7}^{2}\nonumber\\
&& e^{2\phi} = H^{-1},\: B_{0} = H^{-1}-1,\nonumber\\
&& H = 1 +
\frac{r_{0}^{6}sinh^{2}\alpha}{r^{6}},\:f=1-\frac{r_{0}^{6}}{r^{6}}
\label{Q1}
\end{eqnarray}
  From this metric, the mass density along
the z direction can be found \cite{m8}:

\begin{eqnarray}
&& \frac{dM_{F1}}{dz} = T_{F1}k +
\frac{16(T_{F1}k\pi)^{3/2}T^{3}}{81T_{D3}}+
\frac{40T_{F1}^{2}k^{2}\pi^{3}T^{6}}{729T_{D3}^{2}}\label{Q2}
\end{eqnarray}

On the other hand, for finite temperature BIon, the metric is
\cite{m7}:
\begin{eqnarray}
&& ds^{2} = -dt^{2} + dr^{2} + r^{2}(d\theta^{2} + sin^{2}\theta
d\phi^{2}) + \sum_{i=1}^{6}dx_{i}^{2}. \label{Q3}
\end{eqnarray}
 Choosing the world volume coordinates of the D3-brane as
$\lbrace\sigma^{a}, a=0..3\rbrace$ and defining $\tau =
\sigma^{0},\,\sigma=\sigma^{1}$, the coordinates of BIon are given
by \cite{m6,m7}:
\begin{eqnarray}
t(\sigma^{a}) =
\tau,\,r(\sigma^{a})=\sigma,\,x_{1}(\sigma^{a})=l(\sigma),\,\theta(\sigma^{a})=\sigma^{2},\,\phi(\sigma^{a})=\sigma^{3}
\label{Q4}
\end{eqnarray}
and the remaining coordinates $x_{i=2,..6}$ are constant. The
embedding function $l(\sigma)$  describes the bending of the
brane. Let $l$ be a transverse coordinate to the branes and $\sigma$
be the radius on the world-volume.  The induced metric on the
brane is then:
\begin{eqnarray}
\gamma_{ab}d\sigma^{a}d\sigma^{b} = -d\tau^{2} + (1 +
l'(\sigma)^{2})d\sigma^{2} + \sigma^{2}(d\theta^{2} +
sin^{2}\theta d\phi^{2}) \label{Q5}
\end{eqnarray}
so that the spatial volume element is $dV_{3}=\sqrt{1 +
l'(\sigma)^{2}}\sigma^{2}d\Omega_{2}$. We impose the two boundary
conditions that $l(\sigma)\rightarrow 0$ for $\sigma\rightarrow
\infty$ and $z'(\sigma)\rightarrow -\infty$ for $\sigma\rightarrow
\sigma_{0}$, where $\sigma_{0}$ is the minimal two-sphere radius
of the configuration. For this BIon, the mass density along the z
direction can be obtained \cite{m7}:
\begin{eqnarray}
&& \frac{dM_{BIon}}{dl} = T_{F1}k + \frac{3\pi T_{F1}^{2}k^{2}
T^{4}}{32T_{D3}^{2}\sigma_{0}^{2}}+
 \frac{7\pi^{2} T_{F1}^{3}k^{3} T^{8}}{512T_{D3}^{4}\sigma_{0}^{4}}\label{Q6}
\end{eqnarray}
Comparing the mass densities for BIon  to the mass density for the
F-strings, we see that the thermal BIon configuration behaves like
k F-strings at $\sigma = \sigma_{0}$. At the corresponding point,
$\sigma_{0}$ has the following dependence on the
temperature:
\begin{eqnarray}
&& \sigma_{0} =
(\frac{\sqrt{kT_{F1}}}{T_{D3}})^{1/2}\sqrt{T}\left[C_{0} +
C_{1}\frac{\sqrt{kT_{F1}}}{T_{D3}}T^{3}\right]\label{Q7}
\end{eqnarray}
where $T_{F1} = 4k\pi^{2}T_{D3}g_{s}l_{s}^{2}$, $C_{0}$, $C_{1}$,
$F_{0}$, $F_{1}$ and $F_{2}$ are numerical coefficients which can
be determined by requiring that the $T^{3}$ and $T^{6}$ terms in
Eqs. (\ref{Q2}) and (\ref{Q6}) agree. At this point, two universes
are born, however the wormhole is not formed yet. The metric of
these Friedman-Robertson-Walker (FRW) universes are:
\begin{eqnarray}
&& ds^{2}_{Uni1} = ds^{2}_{Uni2} = -dt^{2} + a(t)^{2}(dx^{2} +
dy^{2} + dl^{2}), \label{Q8}
\end{eqnarray}
 The mass
density of black F-string, BIon and two  universes should be equal
at the corresponding point:
\begin{eqnarray}
&& \rho_{uni1} + \rho_{uni2} = \frac{dM_{F1}}{dl} =
\frac{dM_{BIon}}{dl} \rightarrow
\nonumber\\
&& 6 H^{2}  = T_{F1}k +
\frac{16(T_{F1}k\pi)^{3/2}T^{3}}{81T_{D3}}+
\frac{40T_{F1}^{2}k^{2}\pi^{3}T^{6}}{729T_{D3}^{2}}\label{Q9}
\end{eqnarray}
where $H$ is the Hubble parameter and has a following functional form with
redshift \cite{m5}:
\begin{eqnarray}
&& H(z)= H_{0}\sqrt{\Omega_{m}(1+z)^{3}+log( exp(1-\Omega_{m}) +
\delta z)}\label{Q10}
\end{eqnarray}
 where $\Omega_{m}$ and $\delta$ are constants. Solving Eqs. (\ref{Q9}) and using (\ref{Q10}), we
obtain the cosmological redshift in terms of temperature:
\begin{eqnarray}
&& z \simeq \frac{1}{ H_{0}^{1/3}}\left(T_{F1}k +
\frac{16(T_{F1}k\pi)^{3/2}T^{3}}{81T_{D3}}+
\frac{40T_{F1}^{2}k^{2}\pi^{3}T^{6}}{729T_{D3}^{2}}+\Omega_{m}-1\right)^{1/3}-1
\label{Q11}
\end{eqnarray}
At the beginning of time, it is $T = \infty\rightarrow z=\infty$. On the
other hand, Eq. (\ref{Q11}) shows that cosmological redshift
depends on the temperature and decreases very fast.

After a short period of time, wormhole is formed between brane
and antibrane due to the F-string charge and  the universe  enters a
phase of inflation. Putting $k$ units of F-string charge along the
radial direction and using Eq.(\ref{Q5}), we obtain
\cite{m6,m7}:
\begin{eqnarray}
l(\sigma)= \int_{\sigma}^{\infty}
d\acute{\sigma}\left(\frac{F(\acute{\sigma})^{2}}{F(\sigma_{0})^{2}}-1\right)^{-\frac{1}{2}}
\label{Q12}
\end{eqnarray}
At finite temperature, BIon $F(\sigma)$ is given by
\begin{eqnarray}
F(\sigma) = \sigma^{2}\frac{4cosh^{2}\alpha - 3}{cosh^{4}\alpha}
\label{Q13}
\end{eqnarray}
where $cosh\alpha$ is determined by the following function:
\begin{eqnarray}
cosh^{2}\alpha = \frac{3}{2}\frac{cos\frac{\delta}{3} +
\sqrt{3}sin\frac{\delta}{3}}{cos\delta} \label{Q14}
\end{eqnarray}
with the definitions:
\begin{eqnarray}
cos\delta \equiv \overline{T}^{4}\sqrt{1 +
\frac{k^{2}}{\sigma^{4}}},\, \overline{T} \equiv
\left(\frac{9\pi^{2}N}{4\sqrt{3}T_{D_{3}}}\right)T, \, \kappa \equiv \frac{k
T_{F1}}{4\pi T_{D_{3}}} \label{Q15}
\end{eqnarray}
In the above equation, $T$ is the finite temperature of BIon, $N$ is the
number of D3-branes and $T_{D_{3}}$ and $T_{F1}$ are the tensions of
brane and fundamental strings respectively. Attaching a mirror
solution to Eq. (\ref{Q12}), we construct a wormhole configuration.
The separation distance $\Delta = 2l(\sigma_{0})$ is between the $N$
D3-branes and $N$ anti D3-branes for a given brane-antibrane
wormhole configuration defined by the four parameters $N$, $k$, $T$ and
$\sigma_{0}$. We have:
\begin{eqnarray}
\Delta = 2l(\sigma_{0})= 2\int_{\sigma_{0}}^{\infty}
d\acute{\sigma}\left(\frac{F(\acute{\sigma})^{2}}{F(\sigma_{0})^{2}}-1\right)^{-\frac{1}{2}}
\label{Q16}
\end{eqnarray}
In in the limit of small temperatures, we obtain:
\begin{eqnarray}
\Delta = \frac{2\sqrt{\pi}\Gamma(5/4)}{\Gamma(3/4)}\sigma_{0}\left(1 +
\frac{8}{27}\frac{k^{2}}{\sigma_{0}^{4}}\overline{T}^{8}\right)\,.
\label{Q17}
\end{eqnarray}
Let us now discuss the non-phantom  inflationary model  in thermal BIon. For this, we
need to compute the contribution of the BIonic system to the 4D-dimensional universe energy momentum tensor. The electromagnetic  tensor for a
BIonic system with $N$  D3-branes and $k$ F-string charges is
\cite{m7},
 \begin{eqnarray}
&& T^{00}=\frac{2T_{D3}^{2}}{\pi
T^{4}}\frac{F(\sigma)}{\sqrt{F^{2}(\sigma)-F^{2}(\sigma_{0})}}\sigma^{2}\frac{4cosh^{2}\alpha
+ 1}{cosh^{4}\alpha} \nonumber \\&& T^{ii}=
-\gamma^{ii}\frac{8T_{D3}^{2}}{\pi
T^{4}}\frac{F(\sigma)}{\sqrt{F^{2}(\sigma)-F^{2}(\sigma_{0})}}\sigma^{2}\frac{1}{cosh^{2}\alpha},\,i=1,2,3
\nonumber \\&&T^{44}=\frac{2T_{D3}^{2}}{\pi
T^{4}}\frac{F(\sigma)}{F(\sigma_{0})}\sigma^{2}\frac{4cosh^{2}\alpha
+ 1}{cosh^{4}\alpha} \label{Q18}
\end{eqnarray}
 These equations show that with increasing temperature
in BIonic system, the energy-momentum tensors decreases. This is
because when spikes of  branes and antibranes are well
separated, wormhole is not formed and there is  no channel for
flowing energy from universe branes into extra dimensions and
temperature is very high.  However when two universe branes are
close each other and connected by a wormhole, temperature
reduced to lower values.

Now, we can discuss the  $f(R)$  gravity model at finite temperature BIon system
and obtain the explicit form of temperature and equation of state
parameter $\omega$. To this end, we use  the approach in
Ref.\cite{m5} to unify BIonic and $f(R)$ inflation and the three
phases of the universe expansion. The key ingredient is to obtain the redshift transition from decelerated to accelerated behavior and viceversa. In general, the $f(R)$ gravity  has the following action:
\begin{eqnarray}
&& S=\int d^{4}x \sqrt{-g}\{f(R)+L_{m}\}\label{Q19}
\end{eqnarray}
where $L_{m}$ is the matter Lagrangian and $g$ is the determinant of
the metric tensor. Here the energy density $\rho_{curv}$ and the pressure
$p_{curv}$  are \cite{m5}:
\begin{eqnarray}
&& \rho_{curv} = \frac{1}{f'(R)}\left\{\frac{1}{2}[f(R)-R
f(R)]-3H\dot{R}f''(R)\right\},\nonumber
\\&& p_{curv} =-\frac{1}{2}[f(R)-R f(R)]+3H\dot{R}f''(R)
+\ddot{R}f''(R)+\dot{R}[\dot{R}f'''(R)-H f''(R)] \label{Q20}
\end{eqnarray}
where the prime denotes the  derivative with respect to $R$ and dot derivative with respect to time.  Following \cite{m5}, the Ricci scalar R
can be written as a function of $z$:
\begin{eqnarray}
&& R = 6H[(1+z)H_{z}-2H], \label{Q21}
\end{eqnarray}
Also, we have with respect to the redshift $z$:
\begin{eqnarray}
&& f'(R)=R_{z}^{-1}f_{z},\nonumber
\\&& f''(R)=(f_{2z}R_{z}-f_{z}R_{2z})R_{z}^{-3}\nonumber
\\&& f'''(R)=\frac{f_{3z}}{R_{z}^{3}}- \frac{f_{z}R_{3z}+3f_{2z}R_{2z}}{R_{z}^{4}}+3\frac{f_{z}R_{2z}^{2}}{R_{z}^{5}}\label{Q22}
\end{eqnarray}
The ansatz for  $f(z)$  which results in the above  expressions are
\cite{m5}:
\begin{eqnarray}
&& f(z)=f_{0}+\frac{1}{1+z}+f_{1}(1+z)+f_{2}(1+z)^{2}, \label{Q23}
\end{eqnarray}
To evaluate the derivatives of $f(R)$ in terms of the Hubble rate,
we make use of the following identities:
\begin{eqnarray}
&& \dot{R}=-(1+z)H R_{z},\nonumber
\\&& \ddot{R}=(1+z)H[H R_{z} + (1+z)(H_{z}R_{z}+H R_{2z})]\label{Q24}
\end{eqnarray}
For these considerations, we adopt a class  of $f(R)$ gravity models
which are consistent with the Solar System tests \cite{m9}. Then,
using Eq.  (\ref{Q20}), the equation of state parameter is
written as \cite{m5}:
\begin{eqnarray}
&& \omega_{curv} =-1
+\frac{\ddot{R}f''(R)+\dot{R}[\dot{R}f'''(R)-H
f''(R)]}{\frac{1}{2}[f(R)-R f'(R)]-3H\dot{R}f''(R)}\sim\nonumber
\\&&\left\{ \begin{array}{cc}
  -z^{7}  & \text{for $z\rightarrow\infty$}\\
  -1+z^{4} & \text{for moderate values of $z$} \\
  -z^{-3} & \text{for $z\rightarrow 0$} \\
\end{array}\right. \label{Q25}
\end{eqnarray}
On the other hand, using Eq. (\ref{Q18}) and assuming that
higher-dimensional stress-energy tensor  has the following
relation with  energy density and pressure ($ T_i^j = {\mathop{\rm
diag}\nolimits} \left[ {\rho, - p, - p, - p, - \bar{p}, - p, - p,
- p} \right])$, we can obtain the equation of state on the universe
brane of the finite temperature BIon system:
\begin{eqnarray}
&& \omega_{BIon} =-\frac{4 cosh^{2}\alpha}{4cosh^{2}\alpha +1
}(1+l'^{2})\label{Q26}
\end{eqnarray}
We assume that the wormhole is created at $T = T_{0}$ and
$\sigma=\sigma_{0}$ and it vanishes at $T = T_{1}$ and
$\sigma_{0}=0$. In this period of time, we can write: ${\displaystyle \sigma_{0} = \frac{ T-T_{1}}{ T_{0}-T_{1}}\sigma} $. Using this result and Eq.
(\ref{Q12}), we obtain the BIonic equation of state in terms of
temperature:
\begin{eqnarray}
&& \omega_{BIon} =-\frac{4 cosh^{2}\alpha}{4cosh^{2}\alpha +1
}\left(1+\frac{T-T_{1}}{(T_{0}-T)(T_{0}-2T_{1}+T)}\right)\label{Q27}
\end{eqnarray}
This equation indicates that the equation of state parameter is
less than -1 during the inflation era and tends to values larger
than negative ones for $T=T_{1}$. Assuming the BIonic equation of
state parameter equal to equation of state  of $f(R)$ given in  Eq.
(\ref{Q25}) and using Eqs.
(\ref{Q21}-\ref{Q24}), we can estimate the
cosmological redshift in terms of temperature, that is :
\begin{eqnarray}
&& z\sim \frac{\frac{\sqrt{3}}{6}T^{8}
+T^{4}-3\sqrt{3}}{T^{4}}\left(1+\frac{T-T_{1}}{(T_{0}-T)(T_{0}-2T_{1}+T)}\right)^{1/7}\label{Q28}
\end{eqnarray}
As it can be seen from this equation, the cosmological redshift was
infinity at $T=T_{0}$ and reduced to lower values with decreasing
temperature. Using this parameter and Eq.(\ref{Q10}),  we can
calculate the Hubble parameter in terms of temperature:
\begin{eqnarray}
&& H(z) \simeq H_{0}\left\{\Omega_{m}\left(\frac{\frac{\sqrt{3}}{6}T^{8}
+T^{4}-3\sqrt{3}}{T^{4}}\right)^{3}\left(1+\frac{T-T_{1}}{(T_{0}-T)(T_{0}-2T_{1}+T)}\right)^{3/7}+\right.\nonumber
\\ &&\left. log\left[
exp(1-\Omega_{m}) + \delta  \frac{\frac{\sqrt{3}}{6}T^{8}
+T^{4}-3\sqrt{3}}{T^{4}}\left(1+\frac{T-T_{1}}{(T_{0}-T)(T_{0}-2T_{1}+T)}\right)^{1/7}\right]\right\}^{1/2}\label{Q29}
\end{eqnarray}
According to this result, Hubble parameter is infinity at
$T=T_{0}$; however, with decreasing temperature, the universe is accelerated to
very low values in a short period of time. From this point of
view, the behavior of this parameter is the same as we expected
from the  thermal history of the universe.

\section{Stage 2: The cosmological redshift in deceleration era  }\label{o2} In this section, we propose a model that
allows to consider the cosmological redshift in deceleration era
of universe. In this stage, with decreasing temperature and
distance between the  two branes,  the wormhole gives its energy to the 
branes and dies, a tachyon is born and the expansion of two FRW
universes is controlled by the tachyonic potential between
branes and evolves from deceleration to acceleration phase.

To construct a decelerated  model, we consider a set of
D3-$\overline{D3}$-brane pairs in the background  (\ref{Q5}) which
are placed at points $l_{1} = l/2$ and $l_{2} = -l/2$ respectively
so that the separation between the brane and antibrane is $l$. For
the simple case of a single D3-$\overline{D3}$-brane pair with
open string tachyon, the action is \cite{m10}:
 \begin{eqnarray}
&& S=-\tau_{3}\int d^{9}\sigma \sum_{i=1}^{2}
V(TA,l)e^{-\phi}(\sqrt{-det A_{i}})\nonumber \;\,\;\mbox{where}\\&&
(A_{i})_{ab}=(g_{MN}-\frac{TA^{2}l^{2}}{Q}g_{Mz}g_{zN})\partial_{a}x^{M}_{i}\partial_{b}x^{M}_{i}
+F^{i}_{ab}+\frac{1}{2Q}((D_{a}TA)(D_{b}TA)^{\ast}+(D_{a}TA)^{\ast}(D_{b}TA))\nonumber
\\&&
+il(g_{az}+\partial_{a}l_{i}g_{ll})(TA(D_{b}TA)^{\ast}-TA^{\ast}(D_{b}TA))+
il(TA(D_{a}TA)^{\ast}-TA^{\ast}(D_{a}TA))(g_{bz}+\partial_{b}l_{i}g_{ll}).
\label{Q30}
\end{eqnarray}
and
  \begin{eqnarray}
&& Q=1+TA^{2}l^{2}g_{ll}, \nonumber \\&&
D_{a}TA=\partial_{a}TA-i(A_{2,a}-A_{1,a})TA,
V(TA,l)=g_{s}V(TA)\sqrt{Q}, \nonumber \\&& e^{\phi}=g_{s}( 1 +
\frac{R^{4}}{l^{4}} )^{-\frac{1}{2}}, \label{Q31}
\end{eqnarray}
Here $\phi$, $A_{2,a}$ and $F^{i}_{ab}$ are the dilaton field, the gauge
fields and field strengths on the world-volume of the non-BPS
brane, respectively;  $TA$ is the tachyon field, $\tau_{3}$ is the brane tension
and $V (TA)$ is the tachyon potential. The indices $a,b$  denote the
tangent directions of the D-branes, while the indices $M,N$  run over the
background ten-dimensional space-time directions. The Dp-brane and
the anti-Dp-brane are labeled by $i$ = 1 and 2 respectively. Then
the separation between these D-branes is defined by $l_{2} - l_{1}
= l$. Also, in writing the above equations we are using the convention
$2\pi\acute{\alpha}=1$.

 Let us consider now, for the sake of simplicity,   only the $\sigma$
dependence of the tachyon field $TA$  and set the
gauge fields to zero. In this case, the action (\ref{Q30}) in the
region that  $r> R$ and $TA'\sim constant$ simplifies to
  \begin{eqnarray}
S \simeq-\frac{\tau_{3}}{g_{s}}\int dt \int d\sigma \sigma^{2}
V(TA)(\sqrt{D_{1,TA}}+\sqrt{D_{2,TA}}) \label{Q32}
\end{eqnarray}
where $D_{1,TA} = D_{2,TA}\equiv D_{TA}$,
$V_{3}=\frac{4\pi^{2}}{3}$ is the volume of a unit $S^{3}$ and
 \begin{eqnarray}
D_{TA} = 1 + \frac{l'(\sigma)^{2}}{4}+ TA^{2}l^{2} \label{Q33}
\end{eqnarray}
where the 'prime'  denotes a derivative w.r.t. its argument
$\sigma$. A potential which has been used in most of papers is
\cite{m11,m12,m13}:
 \begin{eqnarray}
V(TA)=\frac{\tau_{3}}{cosh\sqrt{\pi}TA} \label{Q34}
\end{eqnarray}
The energy momentum tensor is obtained from action by calculating
its functional derivative w.r.t. the background ten-dimensional
metric $g_{MN}$. The precise relation is $T^{MN} =
\frac{2}{\sqrt{-det g}}\frac{\delta S}{\delta g_{MN}}$. We get,
\cite{m10},
 \begin{eqnarray}
&& T^{00}_{i}=V(TA)\sqrt{D_{TA}},  \nonumber \\&&
T^{zz}_{i}=-V(TA)\frac{1}{\sqrt{D_{TA}}}
\left(TA^{2}l^{2}+\frac{\acute{l}^{2}}{4}\right) \nonumber \\&&
T^{\sigma\sigma}_{i} = -V(TA)\frac{Q}{\sqrt{D_{TA}}} \label{Q35}
\end{eqnarray}
Now, using the above equations, we can obtain the equation of state
parameter as:
\begin{eqnarray}
&& \omega_{brane-antibrane} = -\frac{1+TA^{2}l^{2}}{1 +
\frac{l'^{2}}{4}+ TA^{2}l^{2}}\label{Q36}
\end{eqnarray}
This equation indicates that the equation of state parameter is
 bigger than
-1 in the range of $T_{2}<T<T_{1}$ where $T_{1}$ and $T_{2}$ are
temperatures at the beginning and end of deceleration. We assume
that the velocity of branes is very low during this era and we can 
choose $ TA'\sim \eta $ , ${\displaystyle l'\sim \frac{TA}{(1+TA)^{2}}\sim
\frac{\eta}{(1+TA)^{2}}}$ and ${\displaystyle l\sim
l_{0}\left(1-\frac{TA}{1+TA}+\varepsilon\right)}$. Putting this equation of state  parameter
equal to the equation of state parameter in (\ref{Q25}) (which corresponds to the  $f(R)$
gravity and can be applied for all the three phases) and using
Eqs. (\ref{Q21}- \ref{Q24}), we
obtain the cosmological redshift  in terms of the tachyon:
\begin{eqnarray}
&& z\sim 1
-\left[\frac{(1+TA)^{2}+TA^{2}l_{0}^{2}((1+TA)-TA+\varepsilon(1+TA))^{2}}
{(1+TA)^{2}+TA^{2}l_{0}^{2}((1+TA)-TA+\varepsilon(1+TA))^{2}+\eta^{2}}\right]^{1/4}
\label{Q37}
\end{eqnarray}
Eq.(\ref{Q37} ) shows that cosmological redshift depends
on tachyon and with increasing tachyonic potential reduces to
lower values. Using Eq. (\ref{Q10}), we can write the Hubble
parameter in terms of tachyon, that is 
\begin{eqnarray}
&& H\sim H_{0}\left\{2
-\left[\frac{(1+TA)^{2}+TA^{2}l_{0}^{2}((1+TA)-TA+\varepsilon(1+TA))^{2}}
{(1+TA)^{2}+TA^{2}l_{0}^{2}((1+TA)-TA+\varepsilon(1+TA))^{2}+\eta^{2}}\right]^{1/4}\right\}^{3}
\label{Q38}
\end{eqnarray}
This equation indicates that with increasing tachyon, the Hubble
parameter decreases and, at large values of tachyon,  tends to
$H=H_{0}$.

\section{ Stage 3: The cosmological redshift in late-time
acceleration }\label{o3}

In the previous section, we considered the cosmological redshift in the
condition that tachyon field grows slowly ($TA \sim t \sim
\frac{1}{z}$) and we ignored ${\displaystyle TA'=\frac{\partial TA}{\partial
\sigma}}$ and ${\displaystyle \dot{TA}=\frac{\partial TA}{\partial t}}$ in our
calculations. In this section, we discuss that, with the decreasing of the
separation distance between brane-antibrane, tachyon field grows
very fast, and  $TA'$ and $\dot{TA}$ are not negligible. This dynamics leads to the formation of  another
wormhole. In this stage, the universe evolves  from the 
deceleration phase to the acceleration phase and consequently, late-time
acceleration era of the universe accelerates and ends up in a 
big-rip singularity. In this case, the action (\ref{Q30}) gives
the following Lagrangian $L$:
  \begin{eqnarray}
L \simeq-\frac{\tau_{3}}{g_{s}} \int d\sigma \sigma^{2}
V(TA)(\sqrt{D_{1,TA}}+\sqrt{D_{2,TA}}) \label{Q39}
\end{eqnarray}
where
 \begin{eqnarray}
D_{1,TA} = D_{2,TA}\equiv D_{TA} = 1 + \frac{l'(\sigma)^{2}}{4}+
\dot{TA}^{2} -  TA'^{2} \label{Q40}
\end{eqnarray}
and  we assume that $TA l\ll TA'$. Now, we  study the Hamiltonian
corresponding to the above Lagrangian.
 To derive this,  we need the canonical momentum density ${\displaystyle \Pi =
\frac{\partial L}{\partial \dot{TA}}}$ associated with the tachyon:
 \begin{eqnarray}
\Pi = \frac{V(TA)\dot{TA}}{ \sqrt{1 + \frac{l'(\sigma)^{2}}{4}+
\dot{TA}^{2} -  TA'^{2}}}\label{Q41}
\end{eqnarray}
so that the Hamiltonian can be obtained as:
\begin{eqnarray}
H_{DBI} = 4\pi\int d\sigma  \sigma^{2} \Pi \dot{TA} - L
 \label{Q42}
\end{eqnarray}
By choosing $\dot{TA} = 2 TA'$,  we have:
\begin{eqnarray}
H_{DBI} = 4\pi\int d\sigma \sigma^{2} \left[\Pi
(\dot{TA}-\frac{1}{2}TA')\right] + \frac{1}{2}TA\partial_{\sigma}(\Pi
\sigma^{2}) - L
 \label{Q43}
\end{eqnarray}
In this equation, we have, integrating by parts,
the term proportional to $\dot{TA}$, indicating that  the tachyon can
be studied as a Lagrange multiplier imposing the constraint
$\partial_{\sigma}(\Pi \sigma^{2}V(TA))=0$ on the canonical
momentum. Solving this equation yields:
\begin{eqnarray}
\Pi =\frac{\beta}{4\pi \sigma^{2}}
 \label{Q44}
\end{eqnarray}
where $\beta$ is a constant.  Using Eqs. (\ref{Q44}) in (\ref{Q42}), we
get:
\begin{eqnarray}
&& H_{DBI} = \int d\sigma V(TA)\sqrt{1 + \frac{l'(\sigma)^{2}}{4}
+ \dot{TA}^{2} -  TA'^{2}}F_{DBI} ,  \nonumber \\&&
F_{DBI}=\sigma^{2}\sqrt{1 + \frac{\beta}{\sigma^{2}}}\label{Q45}
\end{eqnarray}
The resulting equation of motion  for $l(\sigma)$, calculating by varying
(\ref{Q45}), is
\begin{eqnarray}
&&\left(\frac{l'F_{DBI}}{4\sqrt{1+
\frac{l'(\sigma)^{2}}{4}}}\right)'=0\label{Q46}
\end{eqnarray}
Solving this equation, we obtain:
\begin{eqnarray}
&&l(\sigma) = 4\int_{\sigma}^{\infty} d\sigma
\left(\frac{F_{DBI}(\sigma)}{F_{DBI}\left(\sigma_{0}\right)}-1\right)^{-\frac{1}{2}}=4\int_{\sigma}^{\infty}
d\sigma' \left(\frac{\sqrt{\sigma_{0}^{4}+\beta^{2}}}{\sqrt{\sigma'^{4}-\sigma_{0}^{4}}}\right)
\label{Q47}
\end{eqnarray}
This solution, for non-zero $\sigma_{0}$,  represents a wormhole
with a finite size throat. However, this solution is not complete,
because, the acceleration of branes is ignored. This acceleration
is due to tachyon potential between branes ($ {\displaystyle a\sim \frac{\partial
V(T)}{\partial \sigma}}$). According to recent investigations
\cite{m14}, each of the accelerated branes and antibranes detects the
Unruh temperature (${\displaystyle T=\frac{\hbar a}{2k_{B}\pi c}}$). We will show
that this system is equivalent to the black brane.

The equations of motion obtained from the action (\ref{Q45})  is:
\begin{eqnarray}
\left(\frac{1}{\sqrt{D_{TA}}}TA'(\sigma) \right)'=\frac{1}{\sqrt{D_{TA}}}
\left[\frac{V'(TA)}{V(TA)}\left(D_{TA}-TA'(\sigma)^{2}\right)\right] \label{Q48}
\end{eqnarray}
We can re-obtain this equation in accelerated fame from the
equation of motion in the flat background of (\ref{Q5}):
\begin{eqnarray}
&& -\frac{\partial^{2} TA}{\partial \tau^{2}} + \frac{\partial^{2}
TA}{\partial \sigma^{2}}=0 \label{Q49}
\end{eqnarray}
By using the following re-parameterizations
\begin{eqnarray}
&& \rho = \frac{\sigma^{2}}{w} ,  \nonumber \\&& w=
\frac{V(TA)\sqrt{D_{TA}}F_{DBI}}{2M_{D3-brane}}\nonumber \\&&
\bar{\tau} = \gamma\int_{0}^{t} d\tau' \frac{w}{\dot{w}} - \gamma
\frac{\sigma^{2}}{2}\label{Q50}
\end{eqnarray}
and doing the  following calculations:
\begin{eqnarray}
\left [\left\{(\frac{\partial \bar{\tau}}{\partial \tau})^{2} -
(\frac{\partial \bar{\tau}}{\partial
\sigma})^{2}\right\}\frac{\partial^{2}}{\partial \tau^{2}}+\left\{(\frac{\partial \rho}{\partial \sigma})^{2} - (\frac{\partial
\rho}{\partial \tau})^{2} \right \}\frac{\partial^{2}}{\partial
\rho^{2}}\right]TA=0\label{Q51}
\end{eqnarray}
we have:
\begin{eqnarray}
&& (-g)^{-1/2}\frac{\partial}{\partial
x_{\mu}}[(-g)^{1/2}g^{\mu\nu}]\frac{\partial}{\partial
x_{\upsilon}}TA=0\label{Q52}
\end{eqnarray}
where $x_{0}=\bar{\tau}$, $x_{1}=\rho$ and the metric elements are
obtained as:
\begin{eqnarray}
&& g^{\bar{\tau}\bar{\tau}}\sim
-\frac{1}{\beta^{2}}(\frac{w'}{w})^{2}\frac{(1-(\frac{w}{w'})^{2}\frac{1}{\sigma^{4}})}{(1+(\frac{w}{w'})^{2}\frac{(1+\gamma^{-2})}{\sigma^{4}})^{1/2}}\nonumber
\\&&g^{\rho\rho}\sim -(g^{\bar{\tau}\bar{\tau}})^{-1}\label{Q53}
\end{eqnarray}
where we have used of previous assumption (${\displaystyle \frac{\partial
TA}{\partial t} = \frac{\partial TA}{\partial \tau}= 2
\frac{\partial TA}{\partial \sigma}}$). 

Now, we can compare these elements with the line elements of a
black D3-brane \cite{m15}:
\begin{eqnarray}
&& ds^{2}= D^{-1/2}\bar{H}^{-1/2}(-f
dt^{2}+dx_{1}^{2})+D^{1/2}\bar{H}^{-1/2}(dx_{2}^{2}+dx_{3}^{2})+D^{-1/2}\bar{H}^{1/2}(f^{-1}
dr^{2}+r^{2}d\Omega_{5})^{2}\nonumber
\\&&\label{Q54}
\end{eqnarray}
where
\begin{eqnarray}
&&f=1-\frac{r_{0}^{4}}{r^{4}},\nonumber
\\&&\bar{H}=1+\frac{r_{0}^{4}}{r^{4}}sinh^{2}\alpha \nonumber
\\&&D^{-1}=cos^{2}\varepsilon + H^{-1}sin^{2}\varepsilon \nonumber
\\&&cos\varepsilon =\frac{1}{\sqrt{1+\frac{\beta^{2}}{\sigma^{4}}}}
\label{Q55}
\end{eqnarray}
Putting the line elements of (\ref{Q55}) equal to line elements of
(\ref{Q53}), we find:
\begin{eqnarray}
&&f=1-\frac{r_{0}^{4}}{r^{4}}\sim
1-\left(\frac{w}{w'}\right)^{2}\frac{1}{\sigma^{4}},\nonumber
\\&&\bar{H}=1+\frac{r_{0}^{4}}{r^{4}}sinh^{2}\alpha \sim 1+\left(\frac{w}{w'}\right)^{2}\frac{(1+\gamma^{-2})}{\sigma^{4}} \nonumber
\\&&D^{-1}=cos^{2}\varepsilon + \bar{H}^{-1}sin^{2}\varepsilon\simeq1\nonumber
\\&&
\Rightarrow r\sim \sigma,r_{0}\sim
\left(\frac{w}{w'}\right)^{1/2},(1+\gamma^{-2})\sim sinh^{2}\alpha
\label{Q56}
\end{eqnarray}
The temperature of BIon is ${\displaystyle T=\frac{1}{\pi r_{0} cosh\alpha}}$, see
\cite{m6} for details. Consequently, the temperature of the brane-antibrane
system can be calculated as:
\begin{eqnarray}
&& T=\frac{1}{\pi r_{0}
cosh\alpha}=\frac{\gamma}{\pi}(\frac{w'}{w})^{1/2}\sim \nonumber
\\&&\frac{\gamma}{\pi}\left(tanh\sqrt{\pi}TA+\frac{l'l''+TA'TA''}{1 + \frac{l'(\sigma)^{2}}{4}+
 TA'^{2}}+ \frac{\frac{\beta}{\sigma^{3}}}{1+\frac{\beta}{\sigma^{2}}}\right) \label{Q57}
\end{eqnarray}
Because $\gamma$ depends
on the temperature, we can write:
\begin{eqnarray}
&&\gamma = \frac{1}{cosh\alpha} \sim
\frac{2cos\delta}{3\sqrt{3}-cos\delta
-\frac{\sqrt{3}}{6}cos^{2}\delta}\sim \nonumber
\\&&\frac{2\bar{T}^{4}\sqrt{1+\frac{\beta^{2}}{\sigma^{4}}}}{3\sqrt{3}-\bar{T}^{4}\sqrt{1+\frac{\beta^{2}}{\sigma^{4}}}
-\frac{\sqrt{3}}{6}\bar{T}^{8}(1+\frac{\beta^{2}}{\sigma^{4}})}
\label{Q58}
\end{eqnarray}
Using Eqs. (\ref{Q57}) and (\ref{Q58}), we can approximate
the explicit form of the temperature as:
\begin{eqnarray}
&&T \sim
\left(\frac{4\sqrt{3}T_{D_{3}}}{9\pi^{2}N}\right)\frac{\sqrt[3]{\pi}}{\sqrt[6]{1+\frac{\beta^{2}}{\sigma^{4}}}}\left(tanh\sqrt{\pi}TA+\frac{l'l''+TA'TA''}{1
+ \frac{l'(\sigma)^{2}}{4}+
 TA'^{2}}+ \frac{\frac{\beta}{\sigma^{3}}}{1+\frac{\beta}{\sigma^{2}}}\right)^{-1/3}
\label{Q59}
\end{eqnarray}
This equation shows that with approaching two branes together and
increasing tachyon, the temperature of the system decreases. This result
is consistent with the  thermal history of  the universe since the  temperature
decreases with time. Now, we want to estimate the dependency of
tachyon on time. To this end, we calculate the energy momentum
tensors and equation of state parameter. Using the electromagnetic tensors for
black D3-brane\cite{m6}, we obtain:
\begin{eqnarray}
&& T^{00} =
\frac{\pi^{2}}{2}T_{D3}^{2}r_{0}^{4}(5+4sinh^{2}\alpha)\sim
\frac{\pi^{2}}{2}T_{D3}^{2}\left(\frac{w}{w'}\right)^{1/2}(9 +
\gamma^{-2})\sim \nonumber
\\&& \frac{\pi^{2}}{2}T_{D3}^{2}\left(tanh\sqrt{\pi}TA+\frac{l'l''+TA'TA''}{1 + \frac{l'(\sigma)^{2}}{4}+
 TA'^{2}}+
 \frac{\frac{\beta}{\sigma^{3}}}{1+\frac{\beta}{\sigma^{2}}}\right)^{-1}\left(9+\frac{2\bar{T}^{4}\sqrt{1+\frac{\beta^{2}}{\sigma^{4}}}}{3\sqrt{3}-\bar{T}^{4}\sqrt{1+\frac{\beta^{2}}{\sigma^{4}}}
-\frac{\sqrt{3}}{6}\bar{T}^{8}(1+\frac{\beta^{2}}{\sigma^{4}})}\right)
\nonumber
\\&& \nonumber
\\&& \nonumber
\\&& T^{ii} =-\gamma^{ii}
\frac{\pi^{2}}{2}T_{D3}^{2}r_{0}^{4}\left(1+4sinh^{2}\alpha\right)\sim
-\left(1+\frac{l'^{2}}{4}\right)\frac{\pi^{2}}{2}T_{D3}^{2}\left(\frac{w}{w'}\right)^{1/2}(5
+ \gamma^{-2})\sim \nonumber
\\&& -\left(1+\frac{l'^{2}}{4}\right)\frac{\pi^{2}}{2}T_{D3}^{2}\left(tanh\sqrt{\pi}TA+\frac{l'l''+TA'TA''}{1 + \frac{l'(\sigma)^{2}}{4}+
 TA'^{2}}+
 \frac{\frac{\beta}{\sigma^{3}}}{1+\frac{\beta}{\sigma^{2}}}\right)^{-1}\times \nonumber
\\&&\left(5+\frac{2\bar{T}^{4}\sqrt{1+\frac{\beta^{2}}{\sigma^{4}}}}{3\sqrt{3}-\bar{T}^{4}\sqrt{1+\frac{\beta^{2}}{\sigma^{4}}}
-\frac{\sqrt{3}}{6}\bar{T}^{8}(1+\frac{\beta^{2}}{\sigma^{4}})}\right)
\label{Q60}
\end{eqnarray}
We assume that the wormhole is created at $T = T_{2}$ and
$\sigma_{0}=0$ and will cause to destruction of universe at $T =
T_{rip}$ and $\sigma_{0}=\sigma$. In this period of time, we can
write: $\sigma_{0} = \frac{ T_{2}-T}{ T_{2} -T_{rip}}\sigma$.
Using this and the tensor ($ T_i^j = {\mathop{\rm
diag}\nolimits} \left[ {\rho, - p, - p, - p, - \bar{p}, - p, - p,
- p,
 } \right])$, we can calculate the equation of state parameter:
\begin{eqnarray}
&& \omega_{BIon} = -\frac{(T_{2}
-T_{rip})(1+\beta^{2}+(T-T_{2})^{2})\left(5+\frac{2\bar{T}^{4}\sqrt{1+\frac{\beta^{2}}{\sigma^{4}}}}{3\sqrt{3}-\bar{T}^{4}\sqrt{1+\frac{\beta^{2}}{\sigma^{4}}}
-\frac{\sqrt{3}}{6}\bar{T}^{8}(1+\frac{\beta^{2}}{\sigma^{4}})}\right)}{(T-T_{rip})(T-2T_{rip}+T_{2})\left(9+\frac{2\bar{T}^{4}\sqrt{1+\frac{\beta^{2}}{\sigma^{4}}}}{3\sqrt{3}-\bar{T}^{4}\sqrt{1+\frac{\beta^{2}}{\sigma^{4}}}
-\frac{\sqrt{3}}{6}\bar{T}^{8}(1+\frac{\beta^{2}}{\sigma^{4}})}\right)}\label{Q61}
\end{eqnarray}
This equation shows that for $\beta^{2}>\frac{5}{4}$, the equation
of state parameter is negative  at the beginning of this era
and less than -1 in the range of $T_{rip}<T<T_{rip}$. Putting this
equation of state  parameter equal to the equation of state parameter in (\ref{Q25}) (which
corresponds to unified $f(R)$ model  and can be applied for all the three
phases), we get:
\begin{eqnarray}
&&
z\sim\left[\frac{(T-T_{rip})(T-2T_{rip}+T_{2})\left(9+\frac{2\bar{T}^{4}\sqrt{1+\frac{\beta^{2}}{\sigma^{4}}}}{3\sqrt{3}-\bar{T}^{4}\sqrt{1+\frac{\beta^{2}}{\sigma^{4}}}
-\frac{\sqrt{3}}{6}\bar{T}^{8}(1+\frac{\beta^{2}}{\sigma^{4}})}\right)}{(T_{2}
-T_{rip})(1+\beta^{2}+(T-T_{2})^{2})\left(5+\frac{2\bar{T}^{4}\sqrt{1+\frac{\beta^{2}}{\sigma^{4}}}}{3\sqrt{3}-\bar{T}^{4}\sqrt{1+\frac{\beta^{2}}{\sigma^{4}}}
-\frac{\sqrt{3}}{6}\bar{T}^{8}(1+\frac{\beta^{2}}{\sigma^{4}})}\right)}\right]^{1/3}\label{Q62}
\end{eqnarray}
and
\begin{eqnarray}
&& H\sim
H_{0}\left[1+\frac{(T-T_{rip})(T-2T_{rip}+T_{2})\left(9+\frac{2\bar{T}^{4}\sqrt{1+\frac{\beta^{2}}{\sigma^{4}}}}{3\sqrt{3}-\bar{T}^{4}\sqrt{1+\frac{\beta^{2}}{\sigma^{4}}}
-\frac{\sqrt{3}}{6}\bar{T}^{8}(1+\frac{\beta^{2}}{\sigma^{4}})}\right)}{(T_{2}
-T_{rip})(1+\beta^{2}+(T-T_{2})^{2})\left(5+\frac{2\bar{T}^{4}\sqrt{1+\frac{\beta^{2}}{\sigma^{4}}}}{3\sqrt{3}-\bar{T}^{4}\sqrt{1+\frac{\beta^{2}}{\sigma^{4}}}
-\frac{\sqrt{3}}{6}\bar{T}^{8}(1+\frac{\beta^{2}}{\sigma^{4}})}\right)}\right]\label{Q63}
\end{eqnarray}

 These equations indicate that the cosmological redshift and the Hubble parameter decrease with temperature and
tend to the values of $\Lambda$CDM model ($z\sim 0$ and $H= H_{0}$) at big
rip singularity. As it can be seen from the cosmological redshift in the 
three stages of universe, this parameter is infinity at the
beginning, reduces very fast in the inflation era, decreases with
lower velocity in the deceleration phase,  reduces with higher rate
at the late-time acceleration and finally  tends to zero at the ripping
time. This result is, in principle,  in agreement with recent observational data.

\section{ Discussion and conclusions}\label{o4}
In previous sections, we construct an $f(R)$ gravity model starting from a   brane-antibrane system.
Such an approach allows to obtain a  unified cosmic history which comprises  inflation, deceleration and acceleration phases.  This model, in principle, can be compared  with cosmological
data and can be used to determine  the ripping time. To
this end, let us  calculate the deceleration parameter in
inflation and late time acceleration epochs.
For this goal, we use  the result obtained in \cite{m5}:
\begin{eqnarray}
&& q = -1 +\frac{(1+z)[3\Omega_{m}(1+z)^{2}+\delta/(\alpha +\delta
z)]}{2[\Omega_{m}(1+z)^{3}+ln(\alpha +\delta z)]}\label{Q64}
\end{eqnarray}
where $\alpha=exp(1-\Omega_{m})$. Using the relation  and
Eqs. (\ref{Q28}) and (\ref{Q62}),  the deceleration
parameter at  inflation and late time acceleration epochs are respectively
\begin{eqnarray}
\begin{array}{cc}
  q\sim -\Omega_{m}\frac{\frac{\sqrt{3}}{6}T^{8}
+T^{4}-3\sqrt{3}}{T^{4}}\left(\frac{T-T_{1}}{(T_{0}-T)(T_{0}-2T_{1}+T)}\right)^{3/7}\left[1+\frac{1}{\alpha+\delta\left(\frac{T-T_{1}}{(T_{0}-T)(T_{0}-2T_{1}+T)}\right)^{1/7}}\right]
& \text{ inflation era, $ z\rightarrow \infty$} \\ \\
 q\sim -(\delta+1)\frac{(T_{2}
-T_{rip})(1+\beta^{2}+(T-T_{2})^{2})}
{(T-T_{rip})(T-2T_{rip}+T_{2})}\left[\alpha
-\frac{2}{(T-T_{rip})(T-2T_{rip}+T_{2})}\right] & \text{acceleration
era, $z\rightarrow 0$}
\end{array}
 \label{Q65}
\end{eqnarray}
It is easy to see that if  we choose $\alpha=2$, $\delta=0.7$, $\Omega_{m}=0.7$, $T_{0}=10^{32}$, $T_{1}=10^{9}$, $T_{2}=10^{4}$ and $T_{rip}=0$,
we find that $q=-0.542$ which leads to $T_{universe}= 3.5$ (in proper temperature units). This result is
compatible with SNeIa data \cite{m16}. Besides,
the deceleration parameter is negative in the range 
$T_{1}<T<T_{0}$ and becomes zero at $T=T_{1}$. This means that the
universe inflates in this period of time. On the other hand, 
$q$ is zero at $T=T_{2}$ and negative again in today
acceleration epoch, tending to $-\infty$ at big rip singularity.

In summary, the present  $f(R)$ gravity model, derived  from  a brane-antibrane
system, unifies  inflation, deceleration and acceleration phases
of expansion history. Specifically,  the cosmological redshift $z$ and the 
Hubble parameter can be obtained in terms of temperature $T$.  These
parameters decrease with reducing temperature and tend to the values
of $\Lambda$CDM ($z=0,$ $H=H_{0}$) at present time. It is possible to  show that, at
transition point,  a BIon system  is formed due to the evolutions of
black fundamental strings at transition point, $z\sim T^{2}$. This
BIon system is a configuration  of the  parallel universe-brane and  anti-universe-brane connected by a wormhole in flat space. With
decreasing temperature, the energy of this wormhole flows into the 
universe branes and leads to inflation. We observe that, at
inflation era, the  cosmological redshift reduces  very fast  ($z\sim
T^{4+1/7}$). After the death of the wormhole, inflation ends,
deceleration epoch starts and cosmological redshift decreases with
lower velocity. By approaching the two universe branes together,
a tachyon is originated; it  grows  and causes the creation of a  new wormhole. Late time acceleration era of the universe begins and,
according to the model predictions,  will end up in big-rip singularity. In
this epoch, the cosmological redshift reduces very  fast  ($z\sim
T^{/3}$) and accelerated to zero. The  model can be compared, in principle, with
 observational data and  cosmological parameters can be obtained in terms of temperature and time.
 In a forthcoming paper, we will discuss the model along the track of Ref.\cite{m5} adopting recent observational  data sets.

\section*{Acknowledgments}
\noindent
A. Sepehri would like to thank  Shahid Bahonar University of Kerman for financial support during
investigation in this work. He also thanks Prof. Harmark for his guidance. S. Capozziello is supported by INFN ({\it iniziative specifiche} QGSKY and TEONGRAV).

 \end{document}